\documentclass{PoS}
\usepackage{amsmath, amsthm, amssymb} 

\def \pt {p_{\rm T}}
\def \et {E_{\rm T}}
\def  \met {\not\!\!\et }

\title{Jets and Missing Transverse Energy Reconstruction with CMS}

\ShortTitle{Jets and Missing Transverse Energy in CMS}

\author{\speaker{Didar Dobur}\thanks{On behalf of the CMS Collaboration}\\
        Istituto Nucleare di Fisica Nazionale(INFN) Pisa, Italy\footnote{Now at the University of Florida, Gainsville, USA}.\\
        E-mail: \email{ddidar@email.cern.ch}}

\abstract{We report on the current simulation studies regarding the reconstruction of 
Jets and Missing Transverse Energy (MET) with the CMS detector at the CERN proton-proton 
LHC accelerator. The performance of various jet algorithms is compared, when using 
calorimeter energy deposits as inputs to the algorithm. The plan for obtaining jet energy 
corrections is outlined and data-driven correction methods are described. Finally, the 
performance of MET reconstruction is summarized.}

\FullConference{2008 Physics at LHC\\
		 September 29 - 4 October 2008\\
		 Split, Croatia}

\begin{document}

\section{Jets in CMS} 
Almost all physics channels of interest at the LHC require good understanding of jet 
clustering by means of jet algorithm performances and optimization of its parameters, as 
well as the detector performances. 
In this section, several studies comparing the performance of different jet algorithms namely, 
Fast-$k_{\rm T}$~\cite{kT}, SISCone~\cite{SisCone}, Midpoint Cone~\cite{Midpoint} and 
Iterative Cone~\cite{jetAlgo} are presented. These jet algorithms can be applied to any given set of 
four-vectors, which allows one to study jets of partons, jets of particles remaining after the hadronization, 
and jets of energy deposited in the detector. A successful jet algorithm should provide a good
correspondence between these three levels. It is desired that a jet algorithm is {\it infrared}
and {\it collinear} safe that is being insensitive to addition of soft particles to the input
list (soft gluon radiation) and being insensitive to having a certain energy distributed to 
two collinear particles instead of one, respectively.
In the experiments, {\it collinear unsafety} is introduced by the finite granularity of the Calorimeter 
and/or any $p_{\rm T}$ threshold applied on the input objects. 

\begin{figure}
\begin{center}
  \includegraphics[scale=0.50]{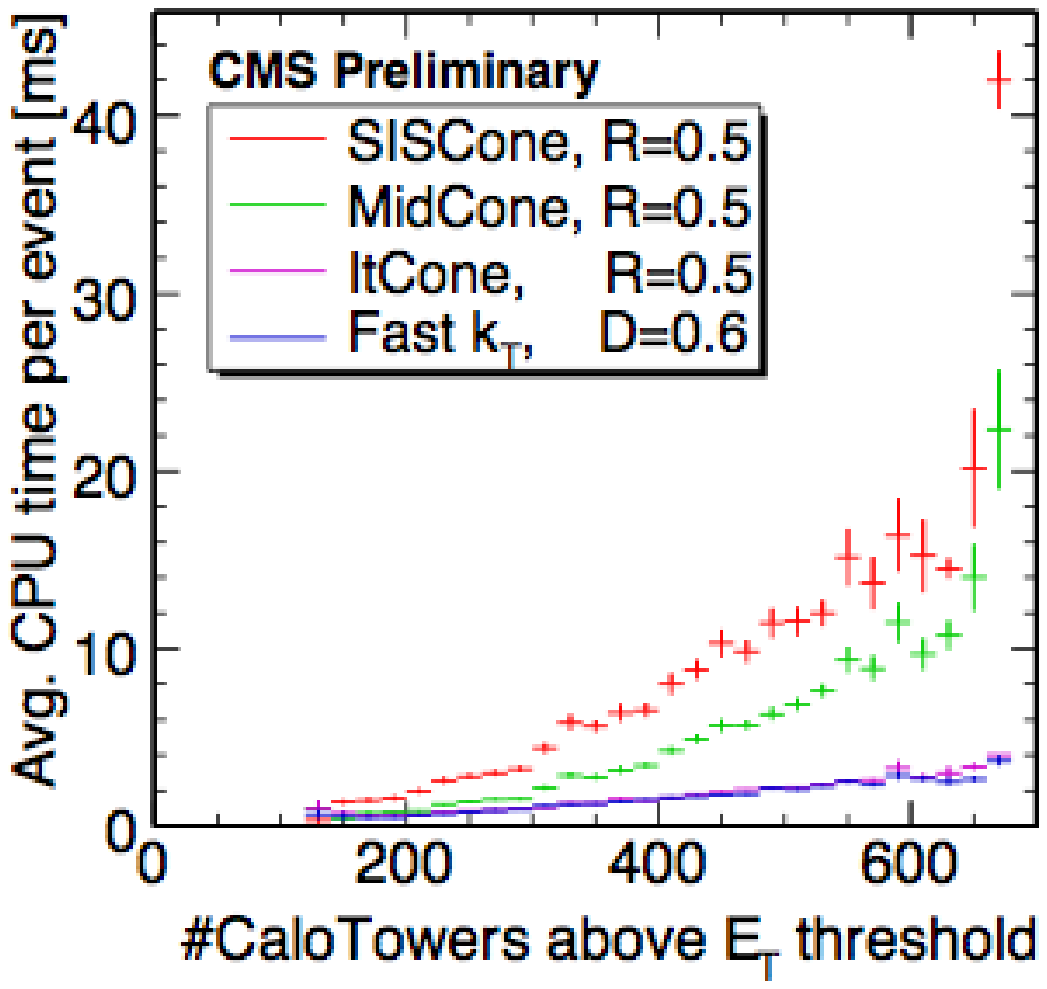}
  \includegraphics[scale=0.60]{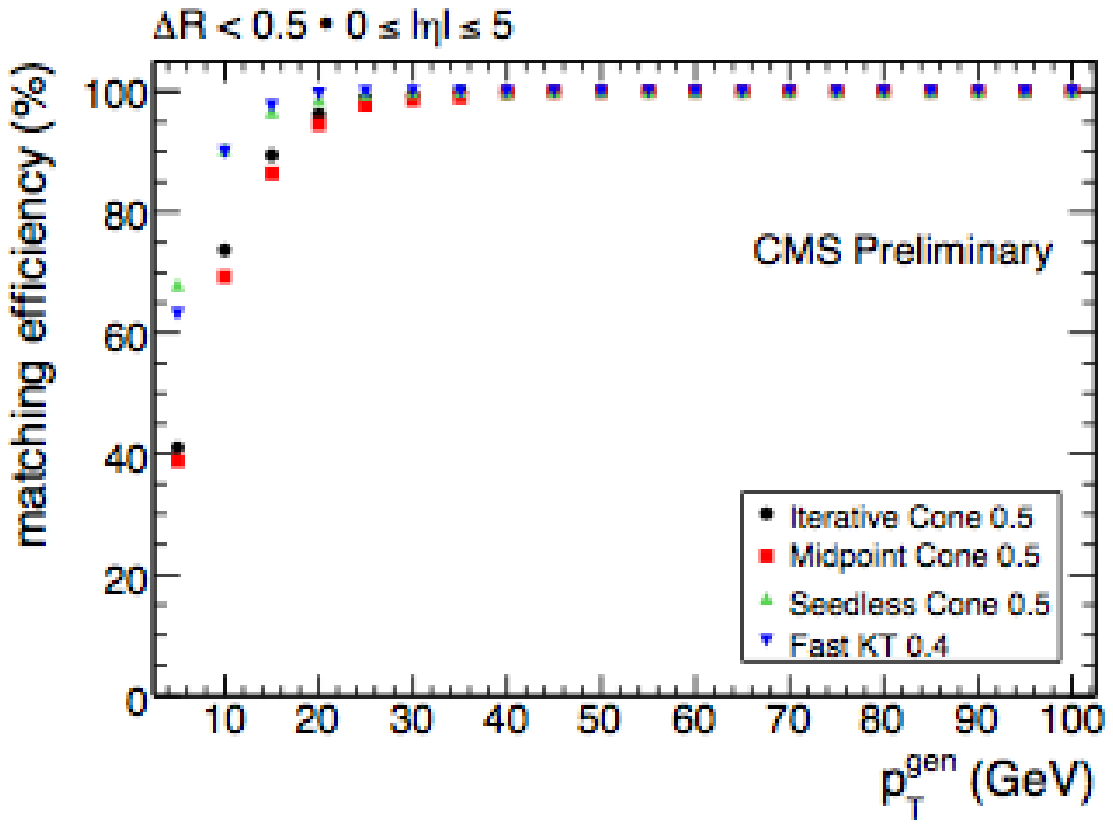} 
  \caption{Left) Average CPU time required for jet reconstruction as a function of the number of CaloTowers
  above $E_{\rm T}$ threshold. Right) Jet Matching Efficiency  versus $p_{\rm T}^{\rm Gen}$. }
  \label{fig:cpu}
\end{center}
\end{figure}

The $k_{\rm T}$ algorithm merges particles by comparing the relative transverse momentum of each pair of
particles in the event and is {\it infrared} and {\it collinear} safe, whereas the 
Cone-based algorithms cluster particles by trying to maximize the energy flow within a cone of radius $R$.
Seedless Infrared Safe Cone (SISCone) algorithm is the latest developed cone-based algorithm which is
{\it infrared safe} to all orders in perturbative expansion and the algorithm itself does not apply a $p_{T}$ 
threshold on the input particles. Another important performance parameter for jet algorithms at hadron
colliders is the execution time. Figure~\ref{fig:cpu} (left) shows the average CPU time per event 
required by each algorithm for jet reconstruction as a function of the number the of the Calorimeter Towers. 
Iterative Cone algorithm is employed by CMS in the High Level Trigger (HLT) for its short and predictable 
execution time. Note that with the {\it fast} implementation of $k_{\rm T}$ algorithm 
the execution time has been reduced significantly. 

In order to quantify the jet reconstruction efficiency, a matching procedure between generator level 
(GenJets) and Calorimeter (CaloJets)
jets is performed. The jet matching efficiency, shown in Fig.~\ref{fig:cpu} (right), is defined as the 
fraction of GenJets which are matched to CaloJets with $\Delta R<0.5$, 
where $\Delta R= \sqrt{\Delta \eta^2 + \Delta \phi ^2}$, $\eta$ and $\phi$ represent the pseudorapidity and
the azimuthal angle respectively.   
The SISCone and the $k_{\rm T}$ algorithms tend to give better efficiencies for very low $p_{\rm T}$ jets while
all algorithms yield $\sim 100 \%$ efficiency for higher $p_{\rm T}$ jets.

\begin{figure}
\begin{center}
  \includegraphics[scale=0.60]{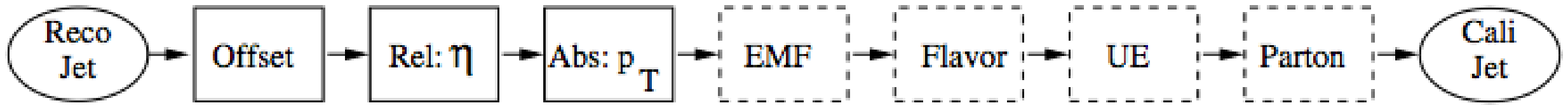}
  \caption{Schematic picture of the factorized multi-level jet correction, where required corrections are
  shown solid boxes and analysis dependent optional ones are shown in dashed boxes.}
  \label{fig:JetCorPlan}
\end{center}
\end{figure}

Jet corrections are necessary for obtaining meaningful jet measurements that are consistent with the jets of
partons emerging from a QCD hard collision. CMS is developing a factorized multi-level jet correction, 
shown schematically in Fig.~\ref{fig:JetCorPlan}, 
where each sub-correction is associated with different detector and physics effects. Most of these corrections can
be obtained from collision data. Initially, however, the corrections will be derived from MC tuned on test beam data.  
   
\begin{figure}
\begin{center}
  \includegraphics[scale=0.55]{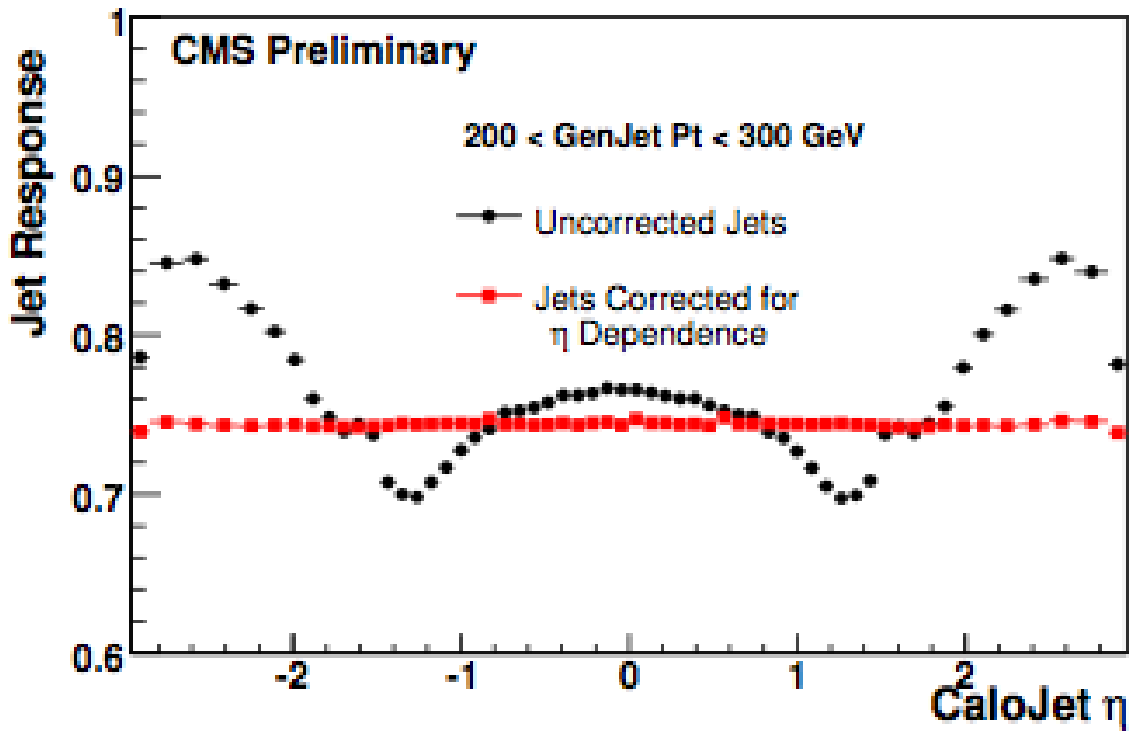}
  \includegraphics[scale=0.55]{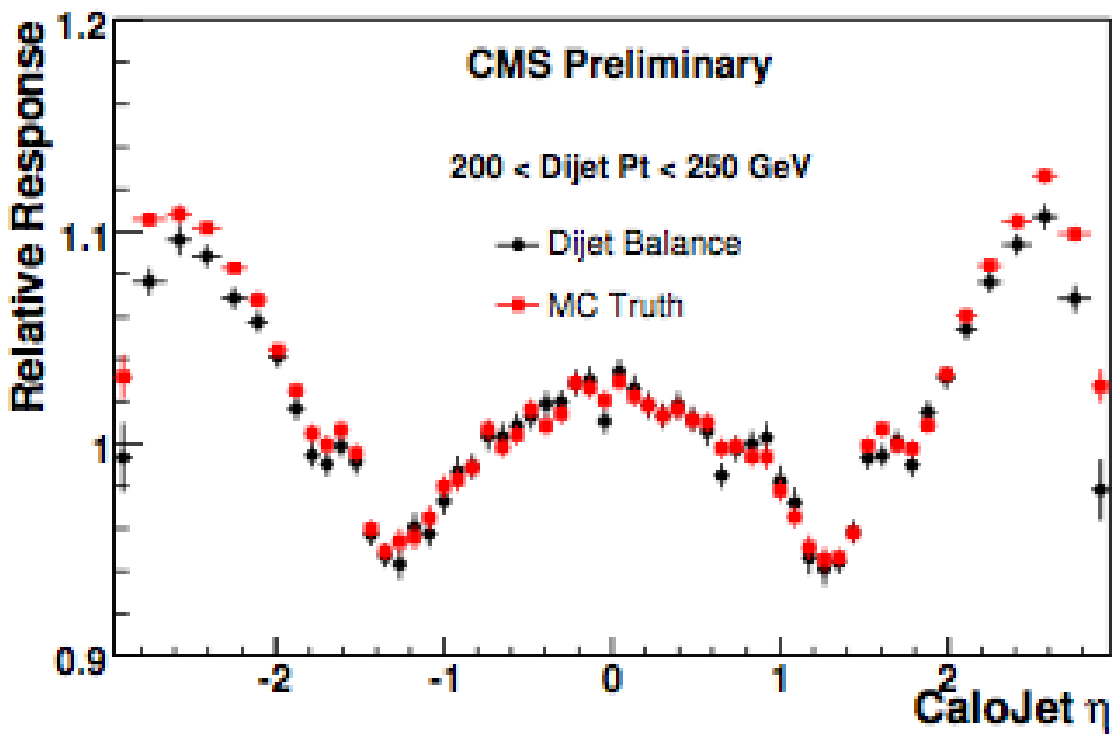} 
  \caption{Left:Jet energy response as a function of pseudorapidity both before and after $\eta$
  dependent corrections. Right:The relative jet response vs. $\eta$ obtained from both di-jet balance
  and MC methods.}
  \label{fig:etaJetRelative}
\end{center}
\end{figure}

The offset corrections are applied to account for pile-up (PU) and the electronic noise in the detector and 
are expected to be relatively small, i.e. on average smaller than 0.5 GeV for a singe pile-up event. 
The non-uniformities of the CMS Calorimeter and its non-linear response to the particles with different
energies necessitate the development of corrections to the measured jet energies as functions of
$|\eta|$ ({\it Relative} $\eta$) and $p_{\rm T}$ ({\it Absolute} $p_{\rm T}$)~\cite{jetcor}. 
{\it Relative}-$\eta$ corrections are
applied in order to achieve a uniform the jet energy response versus $\eta$, where the energy 
of a jet at a given pseudorapidity, is corrected
to the most probable energy in the barrel region ($|\eta|<1.3$) of the Calorimeter. 
The results of these corrections obtained from the MC method are presented in 
Fig.~\ref{fig:etaJetRelative} (left), where significant variations in jet energy response as a function
of $\eta$ is flattened after the corrections. These corrections can be obtained using $p_{\rm T}$ 
balance in back-to-back di-jet QCD events, where one jet being in the barrel region and 
any additional jets are being suppressed by a maximum $p_{\rm T}$ threshold on the additional jets. 
The relative jet energy 
responses obtained using both MC and the data-driven methods are in good agreement 
(see Fig.\ref{fig:etaJetRelative}, right) verifying that such
corrections can be obtained from data.  

\begin{figure}
\begin{center}
  \includegraphics[scale=0.50]{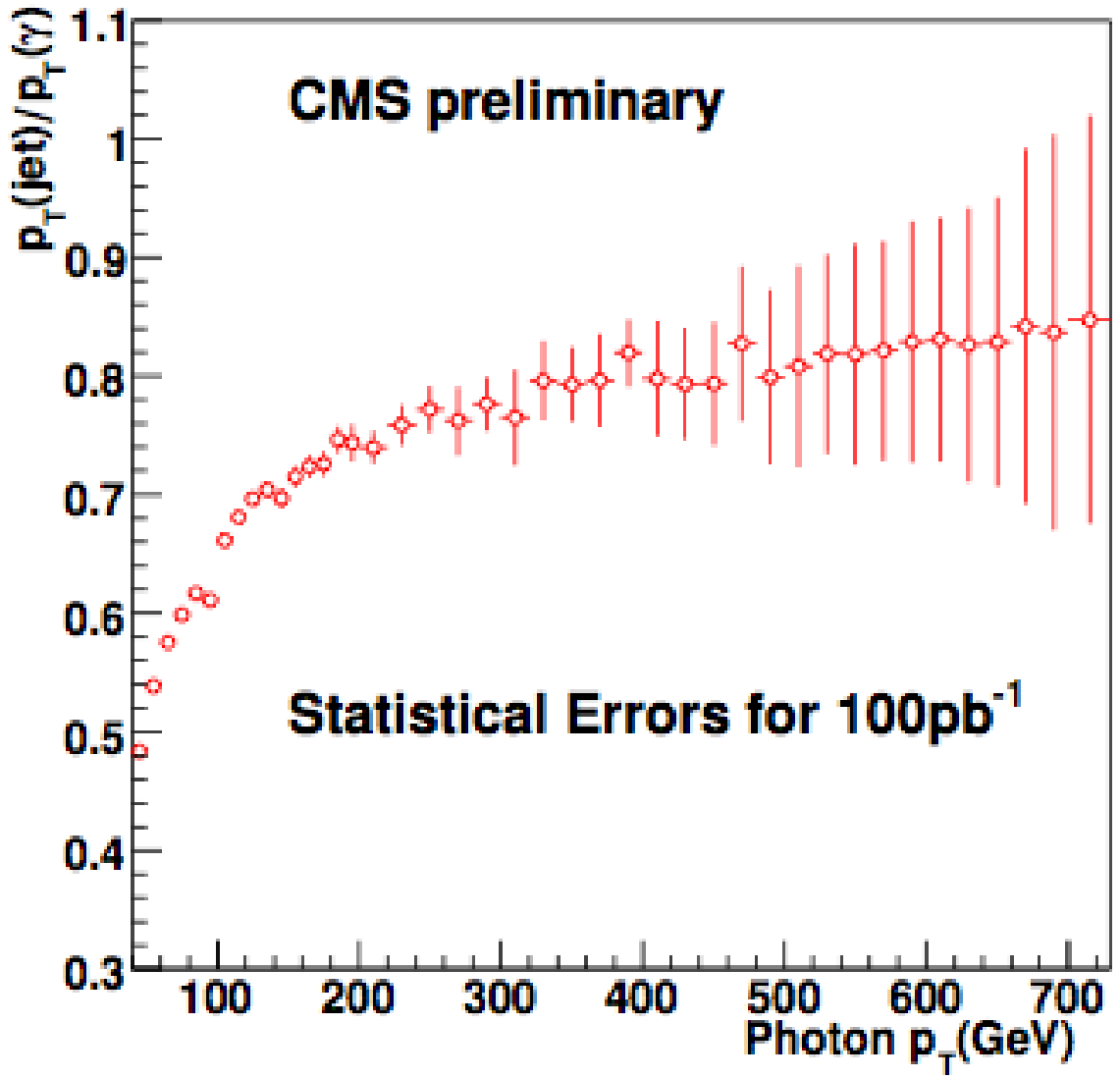}
  \includegraphics[scale=0.55]{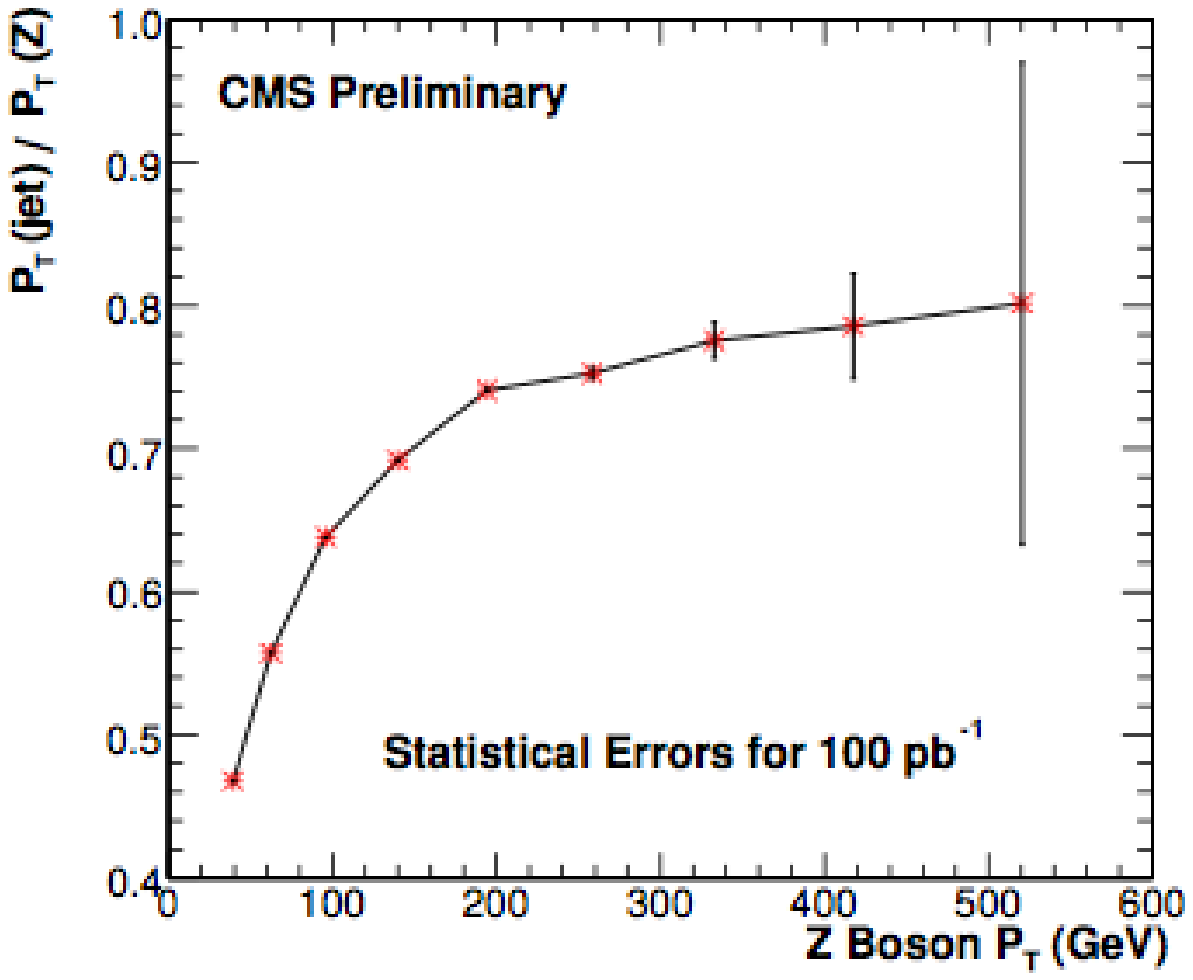} 
  \caption{Jet response as a function of photon(Z) $\pt$ obtained using $\gamma + {\rm Jet}$
  (${\rm Z}+ {\rm Jet}$) balance on the left(right) panel. In both cases the statistical uncertainties 
  are estimated for $100~{\rm pb}^{-1}$.}
  \label{fig:JetCorAbsolute}
\end{center}
\end{figure}

The $p_{\rm T}$ balance in $\gamma/Z+$jet events, where the jet is in the control region ($|\eta|<1.3$), are exploited
for obtaining {\it Absolute} $\pt$ corrections.  
In the case of $\gamma$+jet balance the photon is measured in Electromagnetic Calorimeter (ECAL). The main background
to this process is QCD di-jet events where one jet fakes a photon, and in order to reduce this background
the isolated photons are selected. On the other hand, the Z+jet events, where muon decays of Z are
selected, provide relatively background free channel. 
Muons are identified using the Tracker and the Muon Chambers which give totally independent measurement 
from the Calorimeter where jets are
measured. In both cases, the additional jets are suppressed by an upper threshold on their $p_{\rm T}$. 
Figure~\ref{fig:JetCorAbsolute} shows the expected $p_{\rm T} {\rm (jet)}/p_{\rm T} {\rm (boson)}$ for both 
$\gamma /$Z+jet for $100~pb^{-1}$ of data. Studies show that these corrections can be obtained for jet 
$p_{\rm T}$ up to 400(600) GeV using Z($\gamma$)+jet events. The results are found to be consistent with the ones
obtained with the MC
techniques within $5 \%$, and combining the corrections from different methods by 
extrapolating them to higher $p_{\rm T}$ region is planned. 

The non-compensating feature of the CMS Calorimeter necessitate further corrections which account for the 
differences in 
response depending on the ElectroMagnetic Fraction ({\it EMF}) of the jet energy. These corrections can improve jet energy
resolution up to $10 \%$. On the other hand, the response of the Calorimeter may differ for jets initiated by different
types of partons, i.e. higher response for light quarks. These corrections can reach up to $10 \%$ depending on the
$p_{\rm T}$ region. Further corrections to take into account {\it Underlying Event (UE)} and parton showering 
and hadronization are also
investigated.  

\begin{figure}
\begin{center}
  \includegraphics[scale=0.60]{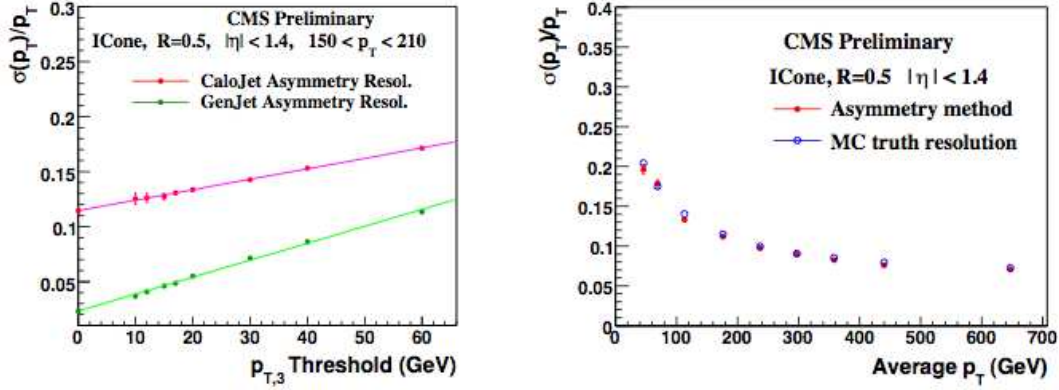} 
  \caption{Left: Jet energy resolution from di-jet asymmetry as a function of the max $p_{\rm T}$ 
  threshold on the $3^{rd}$ jet in the event, extrapolated to zero. Right: Resolutions obtained from Asymmetry Method
  and MC truth.}
  \label{fig:dijetRes}
\end{center}
\end{figure}

CMS develops an {\it Asymmetry Method}, a technique for measuring the jet energy resolutions from data, 
which relates the $p_{\rm T}$ resolution to the resolution of $p_{\rm T}$-imbalance between the two leading jets. By
selecting events with an additional $3^{rd}$ jet and studying the resolution as a function of the max $p_{\rm T}$
requirement on the extra jet, a soft radiation correction is taken into account (see the left pane in
Fig.~\ref{fig:dijetRes}).
The results found to be consistent with the MC based methods verifying that jet energy resolution can be measured
directly from data (see the right pane in Fig.~\ref{fig:dijetRes} right).   
\section{Missing Transverse Energy in CMS}
The large pseudorapidity coverage of the CMS detector allows for rather precise test of 2D-momentum 
conservation in the plain perpendicular to
the direction of the beams. Therefore, any measured significant imbalance in transverse momentum, 
{\it Missing Transverse Energy} ($\met$), can
be considered as the signature of weakly interacting particles which typically escape the detector 
without being measured.      
$\met$ in CMS is determined from the vector sum over uncorrected transverse energy deposits in 
projective Calorimeter Towers~\cite{met}: 
\begin{equation*}\tag{1}
  \label{eq:met}
\vec{\met} = - \sum_{n}(E_{n}\sin \theta_{n} \cos \phi_{n} \hat{i} + E_{n}\sin \theta_{n} \sin \phi_{n} \hat{j}) =
\not\!\!E_{x}\hat{i} + \not\!\!E_{y}\hat{j}.
\end{equation*} 

\begin{figure}
\begin{center}
  \includegraphics[scale=0.55]{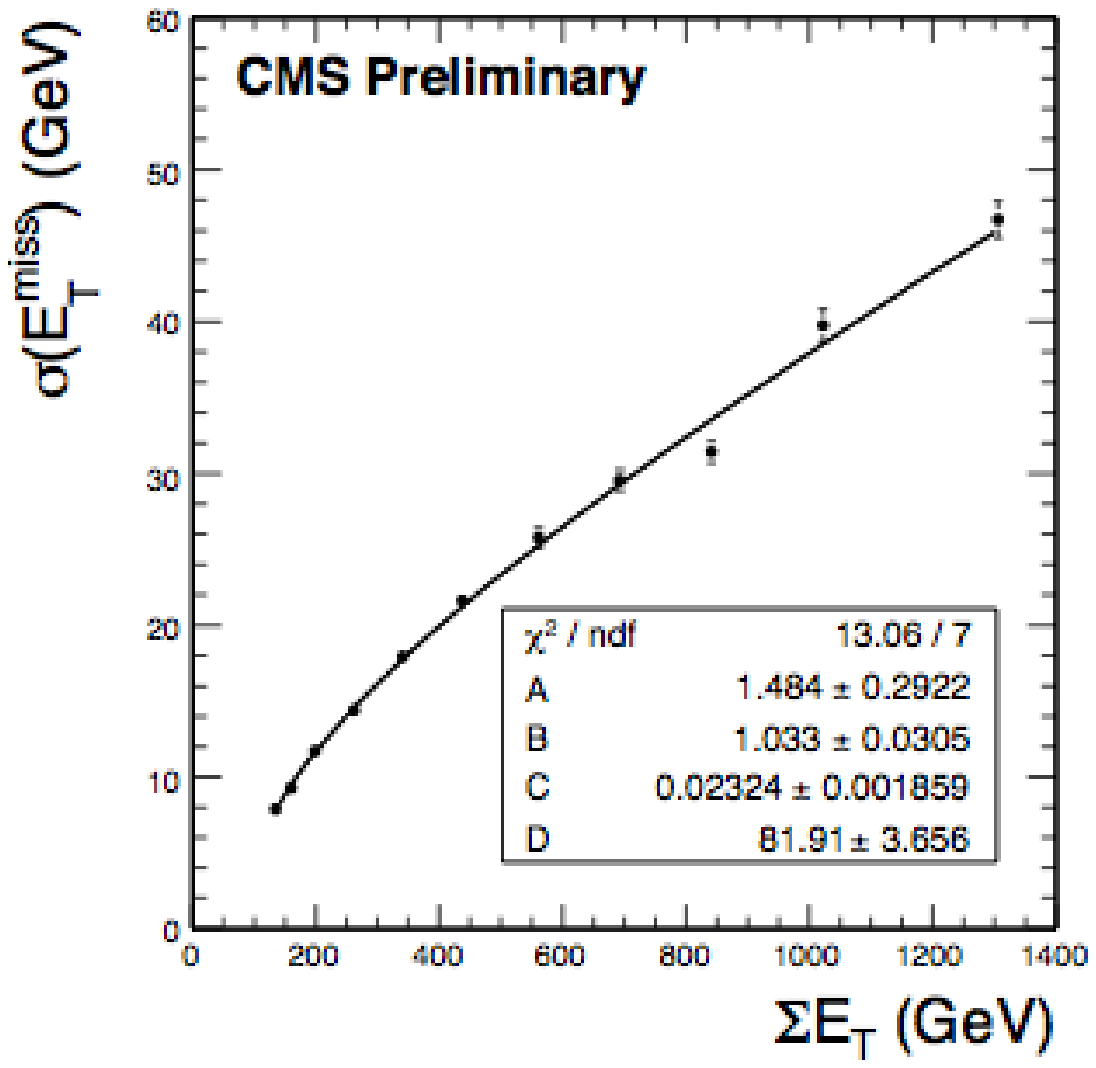}
  \includegraphics[scale=0.65]{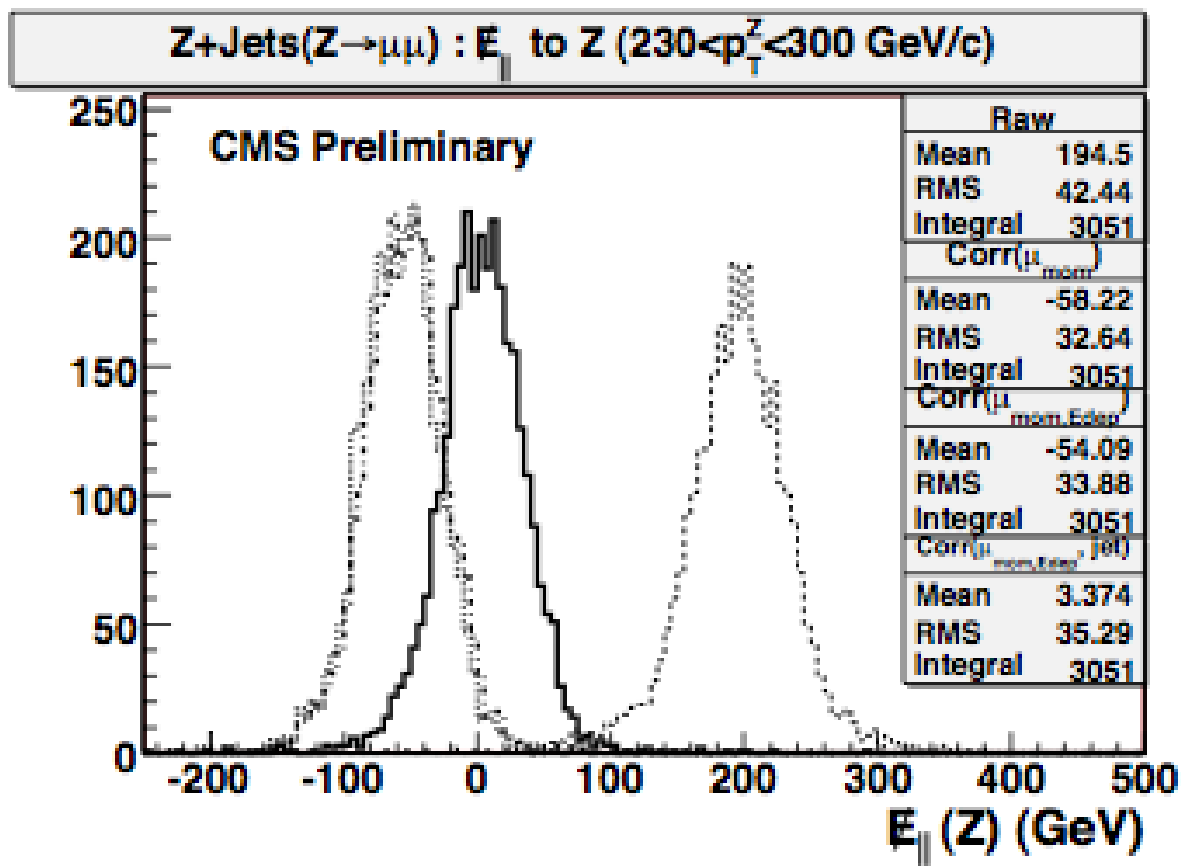} 
  \caption{Left: $\met$ resolution versus $\sum \et$ for QCD dijet sample. Right: The improvements on 
  $\met$ projection to Z boson direction gained
  after applying muon and jet calibration corrections on $\met$.}
  \label{fig:Met}
\end{center}
\end{figure}
$\met$ is a crucial observable for measuring not only the Standard Model (SM) processes (i.e. W boson, top quark production) 
but also for searches for new physics beyond the SM which are associated with relatively large $\met$. However, $\met$ is extremely
sensitive to various detector malfunctions and particles hitting poorly instrumented regions of the detector. Therefore, a great care has
to be taken in order to understand the missing transverse energy distributions in detail before claiming any discovery beyond SM.   

Figure~\ref{fig:Met} (left) shows the missing transverse energy resolution as a function of scalar sum of the
transverse energies in the Calorimeter Towers ($\et$) in QCD. The resolution 
is parametrized by $\sigma(\met)=A\oplus B\sqrt{\sum{\et}-D}\oplus C(\sum{\et}-D)$, where the $A$ ("noise") term represents effects due to 
electronic noise, PU and UE; the $B$ ("stochastic") term represents the statistical sampling nature of the energy deposit in each
Calorimeter Towers; the $C$ ("constant") term represents residual due to non-linearities, cracks and dead material; $D$ ("offset")
represents the effects of noise and PU on $\sum \et$.

As for jets, $\met$ needs to be corrected for various effects. The jet energy corrections are used for calibrating the measured $\met$.
The vectorial sum of absolute corrections on jet $\pt$ is subtracted from the measured raw $\met$ given in Eq.~\ref{eq:met} in order to obtain
calibrated $\met$. The EMF dependent corrections are also taken account. The muons, as being minimum ionizing particles, 
deposit very small fraction of their energy in the Calorimeter, and hence mimic $\met$. In Fig.~\ref{fig:Met} (right), 
the improvements due to these corrections on
the missing transverse energy projection parallel to the Z boson direction in Z+jets events, are shown.  
Another correction to $\met$ become relevant when there are jets of tau lepton decays, since the response of the Calorimeter 
to such jets differs from those of arbitrary jets. In order to account for this effect, the measured tau energy is replaced by the energy
measured by the Particle Flow (PF) algorithm which provides rather precise tau measurement. This correction improves $\met$ resolution
significantly.
    
\section{Conclusions}
Some of the highlights of jet and missing transverse energy studies in CMS are presented. The performances of the 
most relevant and recent jet algorithms are exercised. CMS develops a factorized jet energy correction, where the correction factors will
be based on MC techniques at the start-up, as possible as data-driven when collision data is available and in long term
ithey will be driven from
 MC tuned on collision data.
The $\met$ is an important but a complicated observable that requires careful studies to understand the detector and beam effects 
on it. The first collision data will be crucial to understand both jet and $\met$ objects and their calibrations.

\end{document}